\documentclass[reprint,twocolumn,floatfix,amsmath,amssymb,aps,prl,superscriptaddress]{revtex4-1}

\usepackage{graphicx,dcolumn,bm,graphicx,comment,url,XCharter,xspace,xr,cleveref}
\usepackage[dvipsnames]{xcolor}

\usepackage[normalem]{ulem}
\usepackage[bitstream-charter]{mathdesign}

\newcommand{\tn}[1]{\textnormal{#1}}
\newcommand{\lp}{\ell_\tn{p}}
\newcommand{\lpm}{\ell^\tn{m}_\tn{p}}

\newcommand{\editRevOne}[1]{{#1}}
\newcommand{\edit}[1]{{#1}}
\newcommand{\tangent}{\hat{\tau}}
\newcommand{\uv}{\hat{\upsilon}}
\newcommand{\normal}{\hat{n}}
\newcommand{\binormal}{\hat{b}}

\newcommand{\slope}{\text{slope}\xspace}

\begin{document}
\title{Spool-nematic ordering of dsDNA and dsRNA under confinement}
\author{James D. Farrell}
\email{farrelljd@iphy.ac.cn}
\affiliation{CAS Key Laboratory of Soft Matter Physics, Institute of Physics, Chinese Academy of Sciences, Beijing, 100190, China}
\affiliation{School of Physical Sciences, University of Chinese Academy of Sciences, Beijing, 100049, China}
\affiliation{Songshan Lake Materials Laboratory, Dongguan, Guangdong 523808, China.}
\author{Jure Dobnikar}
\email{jd489@cam.ac.uk} 
\affiliation{CAS Key Laboratory of Soft Matter Physics, Institute of Physics, Chinese Academy of Sciences, Beijing, 100190, China}
\affiliation{School of Physical Sciences, University of Chinese Academy of Sciences, Beijing, 100049, China}
\affiliation{Wenzhou Institute of the University of Chinese Academy of Sciences, Wenzhou, Zhejiang 325011, China}
\author{Rudolf Podgornik}
\email{podgornikrudolf@ucas.ac.cn}
\affiliation{School of Physical Sciences, University of Chinese Academy of Sciences, Beijing, 100049, China}
\affiliation{Kavli Institute for Theoretical Sciences, University of Chinese Academy of Sciences, Beijing, 100049, China}
\affiliation{CAS Key Laboratory of Soft Matter Physics, Institute of Physics, Chinese Academy of Sciences, Beijing, 100190, China}
\affiliation{Wenzhou Institute of the University of Chinese Academy of Sciences, Wenzhou, Zhejiang 325011, China}
\author{Tine Curk}
\email{corresponding author: tcurk@jhu.edu}
\affiliation{Department of Materials Science and Engineering, Johns Hopkins University, Baltimore, Maryland 21218}
\date{\today}

\begin{abstract}
The ability of double-stranded DNA or RNA to locally melt and form kinks leads to strong non-linear elasticity effects that qualitatively affect their packing in confined spaces.
Using analytical theory and numerical simulation we show that kink formation entails a mixed spool-nematic ordering of double-stranded DNA or RNA in spherical capsids, consisting of an outer spool domain and an inner, twisted nematic domain.
These findings explain the experimentally observed nematic domains in viral capsids and imply that non-linear elasticity must be considered to predict the configurations and dynamics of double-stranded genomes in viruses, bacterial nucleoids or gene-delivery vehicles. 
The non-linear elastic theory suggests that spool-nematic ordering is a general feature of strongly confined kinkable polymers.
\end{abstract}

\maketitle

Spatial organization of genomes in viral containers~\cite{Sun2010} and its physical principles~\cite{Zandi2020} is an outstanding problem with a long history~\cite{Pollard_1953}, resulting in refined models based on cryo-electron microscopy~\cite{Luque2020}, X-ray scattering techniques~\cite{Khaykelson2020}, thermodynamic osmotic pressure methodology~\cite{Gelbart2009} and single molecule studies~\cite{Smith2011}. 
These experiments display a range of morphologies (for a recent review see Ref. \cite{Comas-Garcia2023}).
In the case of bacteriophages, high-resolution 3D reconstructions indicate inhomogeneous ordering of the encapsulated genome, with the dsRNA/DNA partitioning into ordered~\cite{ilca2019multiple,Leforestier-poly} or  disordered~\cite{Livolant} nanodomains.
Similarly, measurements of DNA conformational dynamics during packaging~\cite{Berndsen}, as well as intermittent ejection dynamics~\cite{Chiaruttini2010} implicate multi-domain structures.  
However, so far the general assumption has been that the equilibrium ground state is a single inverted spool~\cite{Purohit2003,Zandi2020}, or multiple spools~\cite{curk2019}.

Mesoscopic theories~\cite{Klug2005,Siber2008,Grason2,inviro2016,Liang2019} and coarse-grained simulations of double-stranded genome packaging~\cite{kindt,petrov,rapaport2016packaging, Marenduzzo3,curk2019,Marzinek2020} are usually performed using a semiflexible polymer model that assumes linear elasticity with bending modulus $B\approx50\,k_\tn{B}T$nm, where $k_\tn{B}$ is the Boltzmann constant and $T$ the temperature.
However, the realistic bending response of dsDNA and dsRNA is highly non-linear due to local melting, which enables the formation of kinks.
For dsDNA, kinks form beyond a critical bending torque ${\tau_\tn{c}\approx 30}$~pN nm~\cite{yan2004localized,qu2011critical,qu2011complete}, and typically comprise about three melted base pairs with the reversible work required to form a kink ${\mu \approx 12k_\tn{B}T}$~\cite{qu2011critical}.
Such non-linear effects can alter optimal packing configurations; in particular, a spherically-confined polymer that is able to kink exhibits increased local nematic ordering near the sphere surface~\cite{Myers2017,Bores2020}.

A general theory of a semi-flexible kinkable chain postulates an elastic free-energy density per chain length 
\begin{equation}
f = \frac{B}{2}\kappa^2  + \mu \rho_\tn{k}\;,
\label{eq-Eel2}
\end{equation}
with $\kappa$ the local curvature, $\mu$ the free energy per kink and $\rho_\tn{k}$ the local density of kinks in the chain.
Here, we first perform non-equilibrium simulations to understand the packing arrangement of a kinkable, semi-flexible chain actively pushed into a spherical enclosure, and then analyze the stability of different packings using analytical theory.

A coarse-grained model that accurately describes the properties of DNA or dsRNA at small curvatures is the worm-like-chain (WLC) model;
a semi-flexible polymer of length $L$ and circular cross-section $\sigma$ with persistence length $\lp=B/k_\tn{B}T$.
Efficient description of nonlinear elasticity includes the kinks as regions with locally reduced persistence length~$\lpm$.
In the two-state kinkable WLC (KWLC) model~\cite{wiggins2005exact}, molten regions behave as freely-jointed chains ($\lpm=0$), while the more accurate meltable WLC (MWLC)~\cite{yan2004localized,sivak2012consequences} assigns a single-stranded persistence length $\lpm \approx2~\tn{nm}$ to melted sections. 

\begin{figure}
\centering
\includegraphics[width=\columnwidth]{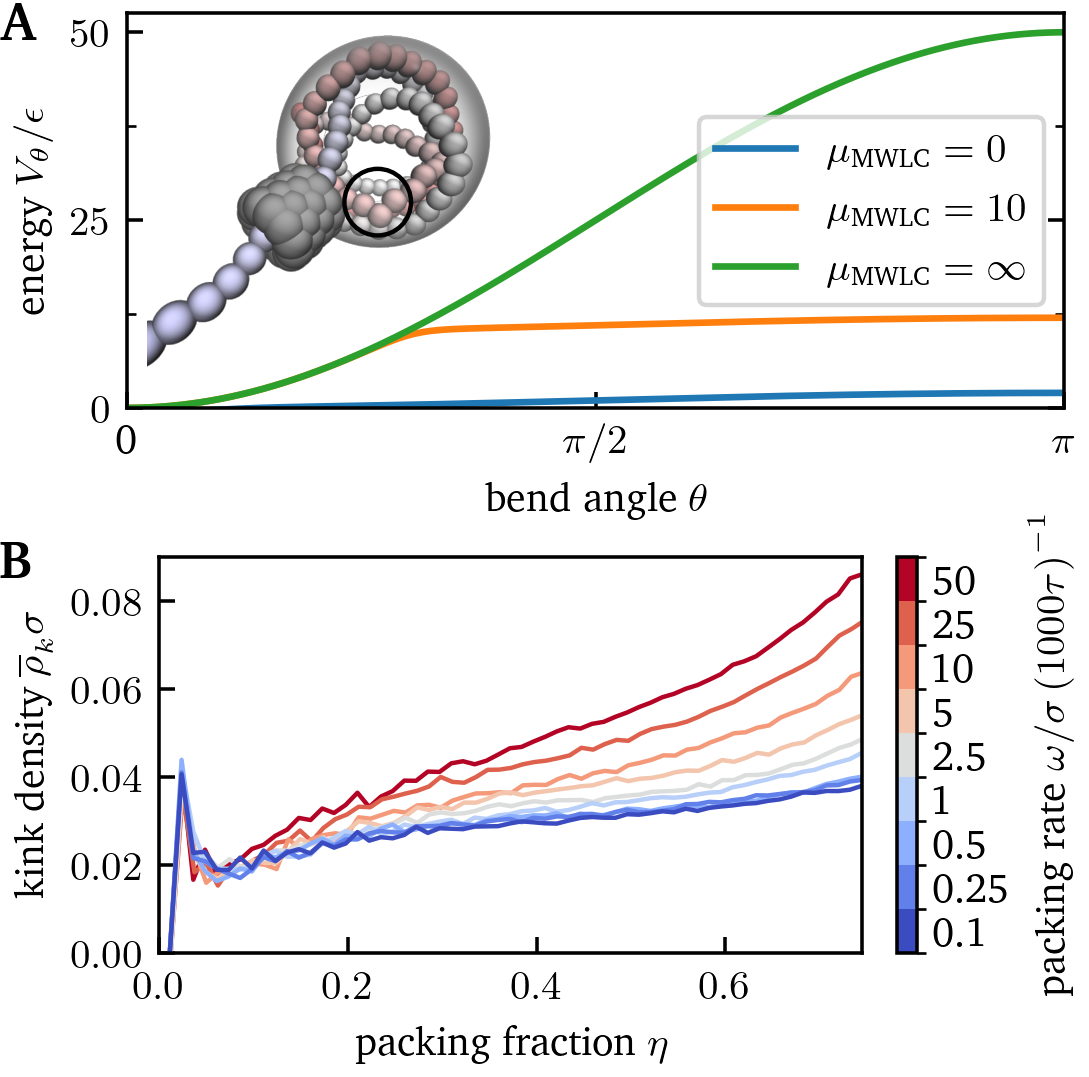}
\caption{
\textbf{A}: Bending potential of the meltable polymer (MWLC) for ssDNA ($\mu_{\tn{MWLC}}=0$), dsDNA ($\mu_{\tn{MWLC}}=10k_\tn{B}T$), and dsDNA without melting ($\mu_{\tn{MWLC}}=\infty$) at $\lp=25\sigma,\,\lpm=1\sigma,\,\sigma=2$~nm.
Inset: Packing simulation setup with the capsid and portal (grey) and partially packed chain (colored by chain index) at $\mu_{\tn{MWLC}}=10k_\tn{B}T$.
The black circle highlights a kink.
\textbf{B}: Corresponding kink density~$\overline{\rho}_\tn{k}$ as a function of packing fraction and packing rate.
}
\label{fig:mwlc}
\end{figure}

We model dsDNA as a discrete MWLC: a bead-spring polymer with $N$ beads of diameter~$\sigma=2\,\tn{nm}$.
Consecutive beads are bonded by a two-body stretching term ${V_{\tn{b}} = K \left(r/\sigma - 1\right)^{2}}$, where $r$ is the distance between two consecutive beads and the prefactor ${K=16k_\tn{B}T \lp/\sigma}$~\cite{curk2019}, while excluded volume \editRevOne{and short range hydration interactions at relevant packing densities \cite{inviro2016} are} incorporated as a WCA repulsion~\cite{weeks1971role} with strength ${\epsilon=k_{\tn{B}}T}$.
To enable molecular dynamics (MD) simulations, we adapt the two-state MWLC model by writing it as a continuous, three--body bending interaction obtained as a canonical-ensemble superposition of a non-melted and a melted state.
Assuming the states in each three-body segment are independent, and defining the Boltzmann factors
${q = \exp{\left[- \lp  (1-\cos{\theta})/\sigma\right]}}$ and
${q^\tn{m} = \exp{\left[- \beta \mu_{\tn{MWLC}} - \lp^{\tn{m}} (1-\cos{\theta})/\sigma\right]}}$,
the potential of mean force is
\begin{equation}
V_{\theta}= -k_\tn{B}T \log \left[ q+q^\tn{m} \right] \;,
\end{equation}
with $\theta$ the angle between consecutive bond vectors, and $\mu_{\tn{MWLC}}$ the melting penalty per bead (Fig.~\ref{fig:mwlc}).
We obtain the average kink density~$\overline{\rho}_\tn{k}$ as the canonical average, $\overline{\rho}_\tn{k}=\left\langle q^\tn{m}/\left(q+q^\tn{m}\right)\right\rangle/\sigma$.
\edit{By fitting to experimental data, we find} the MWLC model captures the non-linear elasticity of DNA bending at parameters $\lpm=\sigma$, $\mu_{\tn{MWLC}}=10 k_\tn{B}T$ and reproduces the kink formation of dsDNA chain at critical torque $\tau_\tn{c}=32\,\tn{pN nm}$ \edit{(see Fig.\,S1 in \cite{Note1}).}
\nocite{qu2010elastic,fiorin2013using}
\editRevOne{Obtaining the kink formation free-energy at a specific base pair requires de-convolution of the coarse-grained model~\cite{schopflin2012probing}, but since $\mu\gg k_\tn{B}T$, the probability that more than one kink occurs within a single bead is negligible.
Thus,} the full ($180^{\circ}$) kink formation penalty at a specific base-pair is \edit{$\mu=\mu_{\tn{MWLC}}{(\sigma)} + 2k_\tn{B}T + k_\tn{B}T\log(\sigma/d_\tn{n}) \approx 13.8k_\tn{B}T$}, where the last term \editRevOne{re-scales} the number of kink states at bead size $\sigma$ relative to length per nucleotide $d_\tn{n}\approx0.33~\tn{nm}$, \edit{and we  emphasize that $\mu_{\tn{MWLC}}$ is a function of the discretization length~$\sigma$}.

Using the MWLC model we perform MD simulations of DNA packing into a spherical capsid~\cite{plimpton1995fast}.
Confinement is implemented as a WCA repulsion at $R=6\sigma$ with strength $\epsilon$ and diameter $\sigma$, giving an effective enclosure radius $R_{\tn{eff}}=5.5\sigma$.
The maximum packing density of chains is determined by hexagonal close-packing~\cite{inviro2016}.
Taking the volume of the chain relative to the maximum packed volume we define the packing fraction \mbox{$\eta\equiv 3\sqrt{3}\sigma^3N_\tn{in}/(8\pi R_{\tn{eff}}^3)$} where $N_\tn{in}$ the number of beads packed in the capsid.

We model the viral packing motor as a cylindrical portal of length $\lambda_\tn{p}=2\sigma$  and radius $R_\tn{p}=1.2\sigma$ 
(Fig.~\ref{fig:mwlc}A inset).
The motor force operates along the portal axis on beads inside the portal, implemented as a parabolically-modulated traveling wave 
\begin{equation}
{F_{\tn{motor}}=\frac{4f_\tn{s}x(\lambda_\tn{p}-x)}{\lambda_\tn{p}^{2}}}  \sin{\left(2\pi\frac{x-\omega t}{\sigma}\right)}\;,
\label{eq:fmotor}
\end{equation}
where $x$ is the bead position in the portal ($0\le x\le\lambda_\tn{p}$), $\omega$ the packing rate, $t$ the time, and $f_\tn{s}=50k_{\tn{B}}T\sigma^{-1}$ the 
\editRevOne{stall} force.
\editRevOne{
Real viral motors pack in a discrete fashion, at a rate dependent on the availability of ATP~\cite{fuller2007single,fuller2007measurements}.
Our periodic motor force mimics this behavior, and also allows us to tune the stall force and packing rate independently.
}
To model diffusive dynamics in an aqueous solution we apply a Langevin thermostat with damping time~${\tau=\sqrt{m \sigma^2 / (k_{\rm B}T)}}$ where $m$ is the mass of the beads, which \textit{via} the Stokes-Einstein relation~\cite{Note2} results in a simulation timescale $\tau \approx 18$~ns.
\edit{Since the torsional strain can relax on the packing timescale~\cite{rollins2008role,rapaport2016packaging}, we omit torsional terms from our models.}

Using this setup we ran packing simulations of a 600-bead chain into a radius $R_{\tn{eff}}=5.5\sigma$ sphere at rates from $\omega=1\times10^{-4}\sigma\tau^{-1}$ to $ 5\times10^{-2}\sigma\tau^{-1}$, with 120 independent simulations per $\omega$ 
\editRevOne{and evaluate the line density of spontaneously formed kinks (Fig.~\ref{fig:mwlc}B).
The initial peak corresponds to the appearance of the first kink during packing. 
The observed kink density is strongly rate-dependent; however, at slowest packing rates,
$\omega \sim 1 \times 10^{-4} \sigma/\tau \approx 40 \,\tn{kb}/\tn{s}$, it converges to $\overline{\rho}_\tn{k} \approx 0.03/\sigma$. 
This rate is still an order of magnitude faster than in real viral motors $\omega \lesssim 2\tn{kb}/\tn{s}$~\cite{fuller2007measurements,fuller2007single}, but the observed convergence suggests that reducing the rates further would not alter the packed configurations. 
Moreover, the equilibrium theory ($\omega\to0$) shown below agrees well with the slow-packing simulations, suggesting that our predictions capture the biologically relevant packing rates. 
}

Spherically-confined dsDNA is thought to arrange into spools~\cite{Purohit2003,curk2019}, but kinking permits nematic ordering~\cite{Myers2017,Bores2020},
or a mixed phase of spools wrapped around a nematic core~\cite{ilca2019multiple}.
To quantify spool and nematic order in our simulation configurations, we compute the order parameter tensor,
\begin{equation}
    Q_{\alpha\beta} = \frac{3}{2}{\left(\frac{1}{N_\tn{in}}\uv_{i\alpha}\uv_{i\beta} - \frac{1}{3}\delta_{\alpha\beta}\right)},
\label{eq:op}
\end{equation}
where $\uv_{i\alpha}$ is the $\alpha^{\tn{th}}$ component of either the unit binormal vector, $\boldsymbol{\binormal}$, or the unit tangent vector, $\boldsymbol{\tangent}$, at chain \mbox{segment $i$}.
The order parameter, $S$, is the principal eigenvalue of $\mathbf{Q}$.

Taking $\boldsymbol{\uv}=\boldsymbol{\binormal}$, gives the spool order parameter, $S_{\tn{spl}}$, similar to that defined in Ref.~\cite{nikoubashman2017semiflexible},
while taking
$\boldsymbol{\uv}=\boldsymbol{\tangent}$ gives the nematic order parameter, $S_{\tn{nem}}$.
\editRevOne{$S_\tn{nem}$ is of limited practical use because the spool domain is always more massive than the nematic domain.
A more useful mixed spool--nematic order parameter $S_{\tn{mix}}$ is defined by setting}
\begin{equation}
\bm{\uv}_{i}=
    \begin{cases}
        \bm{\tangent}_{i} & \tn{for } 0 < r_{i} \le r_{\tn{c}}\\
        \bm{\binormal}_{i} & \tn{for } r_{\tn{c}} < r_{i} < R_{\tn{eff}},\\
    \end{cases}
    \label{eq:smixed}
\end{equation}
where $r_{i}$ is the distance from the center-of-mass of the $i^\tn{th}$ chain segment to a director $\bm{\hat{z}}$, centered at the origin, and $r_{\tn{c}}$ defines the spool--nematic boundary.
In this case, $S_{\tn{mix}}$ must be maximized with respect to both $r_{\tn{c}}$ and $\bm{\hat{z}}$.
\editRevOne{We denote the partial packing fractions of the spool and nematic domains $\eta_{\text{spl}}$ and $\eta_{\text{nem}}$, respectively.}
\begin{figure}[bt!]
\centering
\includegraphics[width=\columnwidth]{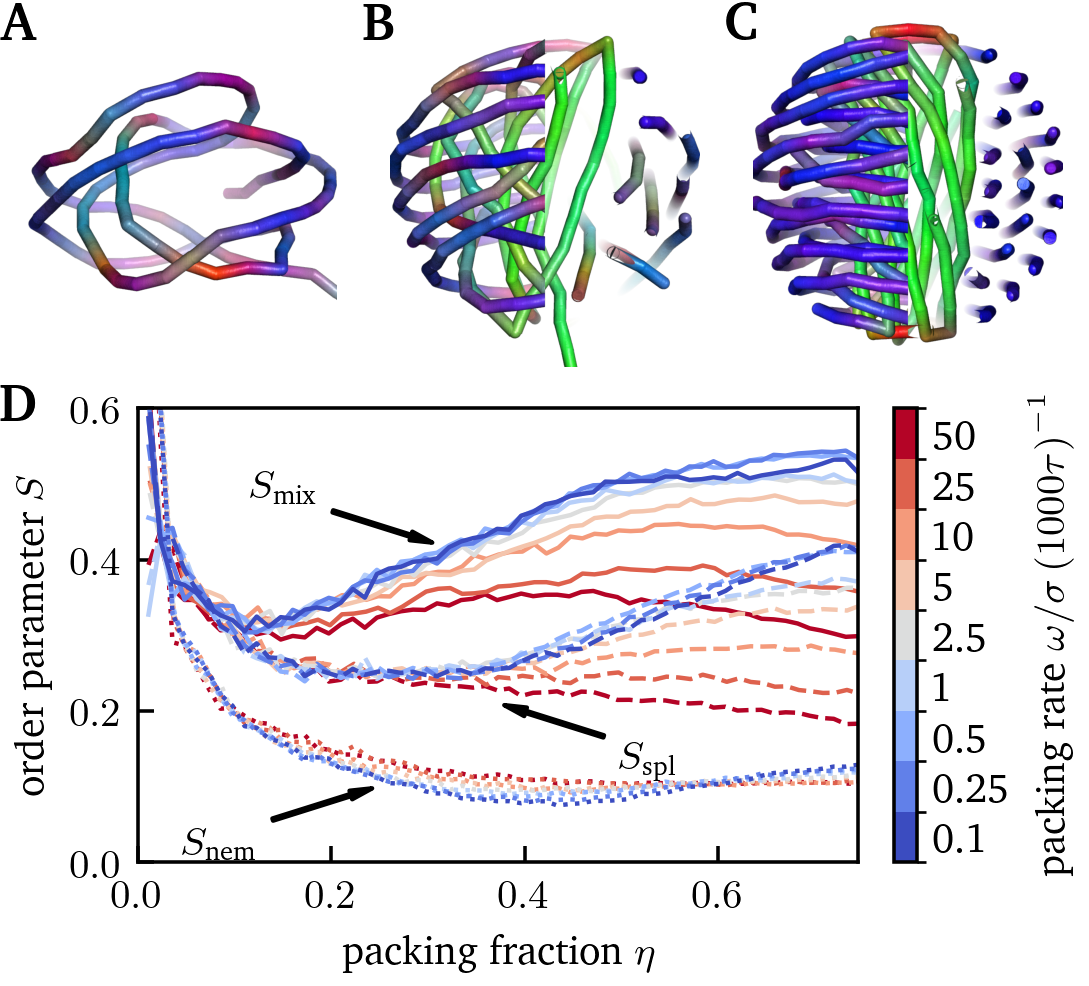}
\caption{\textbf{A--C}: representative configurations at packing fractions $\eta= 0.12,\,0.37,\,0.75$, respectively, of a MWLC packed into a spherical capsid of size $2R_{\tn{eff}}=11\sigma$ at packing rate ${\omega=0.0001\sigma\tau^{-1}}$.
Each segment is assigned an RGB color according to the projection on the cylindrical axis, $\left(\tn{red},\tn{green},\tn{blue}\right)=\left(|\normal_{z}|,|\tangent_{z}|,|\binormal_{z}|\right)$.
On the right hand side in (\textbf{B,C}) only chain segments inside a $3\sigma$-thick slab are shown to emphasize the emergence of a twisted-nematic core.
\textbf{D}: mean values of nematic~$S_{\tn{nem}}$ (dotted lines), spool~$S_{\tn{spl}}$ (dashed lines), and mixed~$S_{\tn{mix}}$ (solid lines) order parameters.
Error bars not drawn for clarity; see Fig~S2 in \edit{\cite{Note1}} for uncertainties.
}
\label{fig:order}
\end{figure}

\editRevOne{At the slowest packing rates, we} found that the chains initially arrange into a loose spool~(Fig.\,\ref{fig:order}A).
At intermediate packing densities ($\eta\approx0.3$), a distinct inner domain emerges exhibiting some nematic order~(Fig.\,\ref{fig:order}B), which is reflected in the growth of $S_{\tn{mix}}$ while $S_{\tn{spl}}$ remains constant (Fig.\,\ref{fig:order}D).
\editRevOne{Upon further packing, $S_{\tn{spl}}$ begins to grow faster than $S_{\tn{mix}}$, indicating that the growth of the nematic core has stopped ($\eta_{\text{nem}}$ levels off at $\approx0.16$), while chain continues to be packed into the outer spool,
and the core shrinks in size from $\approx2.8\sigma$ to $\approx2.1\sigma$ (see Figs. S3 and~S4 in \edit{\cite{Note1}}).
These trends, and the snapshot in Fig.~\ref{fig:order}C, confirm our main observation: for dense packings, the genome adopts a mixed spool--nematic configuration.}

Fast and slow packings are indistinguishable at low densities, but the growth of $S_{\tn{mix}}$ is dramatically reduced at faster packing rates, even falling at high densities (Fig.\,\ref{fig:order}D) indicating kinetically-arrested disorder.
Moreover, for the fastest packing rate $r_\tn{pack}=0.05\sigma\tau^{-1}$, the size of the core region $r_{\tn{c}}$ remains roughly constant $r_\tn{c}\approx2.8\sigma$  during the time the kink density doubles from $0.04$ to $0.08$, indicating that the chain has insufficient time to relax and grow the spool domain as packing proceeds.
At slower packing rates the structures are more ordered and approach the spool--twisted-nematic configuration, while the values of all three order parameters and the kink density are well-converged among the three slowest packing rates.

\begin{figure*}[tb!]
\centering
\includegraphics[width=\textwidth]{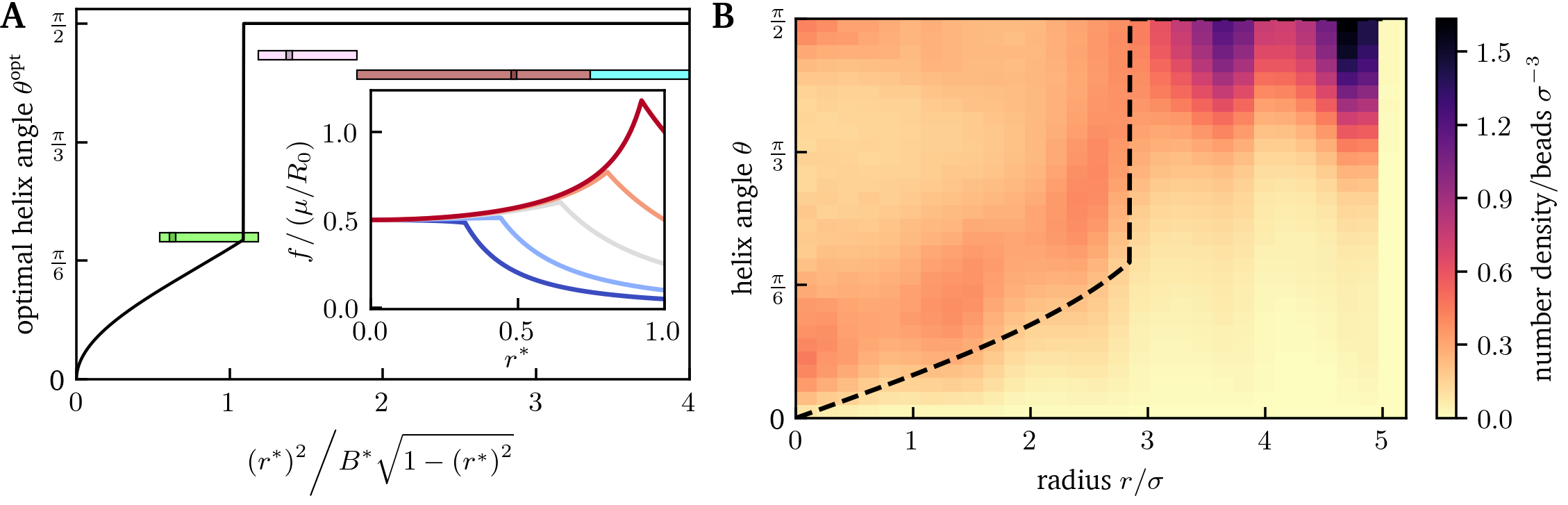}
\caption{\textbf{A}: Optimal helix angle $\theta^\tn{opt}$ as a function of helix radius obtained by minimizing the free-energy density~\eqref{eq-Eel2} (black solid line).
The bars show experimental data for bacteriophage $\phi6$~\cite{ilca2019multiple} (colors chosen to match individual dsRNA shells in Ref.~\cite{ilca2019multiple})
Inset: optimal energy density for ${B^{*}=0.1\, (\tn{blue}),\,0.2,\,0.5,\,1.0,\,2.0\,(\tn{red})}$.
\mbox{\textbf{B}: Distribution} of $\left(r,\theta\right)$ in configurations with packing fractions $\eta=0.59\tn{--}0.62$ from simulations. 
The dashed curve gives the theoretical prediction for $R_{0}/\sigma=5,\,\beta\mu=13.8$.
}
\label{fig:theory-pitch}
\end{figure*}
To investigate whether our packing is under thermodynamic or kinetic control, we consider an analytical theory where 
\editRevOne{a chain confined to a radius $R_{0}$ sphere adopts a helical configuration parametrized as ${u\mapsto\left(r\cos{u},r\sin{u},bu\right), u\in\left[-h/b, h/b\right]}$ with radius $r$, pitch length $2\pi b$, and helix height $h=\sqrt{R_{0}^{2}-r^{2}}$.
The \slope $p=b/r$ corresponds to the helix angle $\theta=\arctan{\left(1/p\right)}$.
The limit $p\to \infty$ corresponds to a nematic, and $p\to0$ to a spool phase. 
To obtain the optimal \slope, we minimize the free energy [Eq.\,~\eqref{eq-Eel2}] assuming that each radial shell is independent of its neighbors.
The local curvature of a helix is ${\kappa(r,b)=|r|/\left(r^{2}+b^{2}\right)}$,
and the contour length of a helix is $L{\left(r,b\right)}=\left(2h/b\right)\sqrt{r^{2}+b^{2}}$.
Assuming a $180^{\circ}$ kink forms whenever the chain hits the confining capsid, the kink density at radius $r$ is $\rho_\tn{k}{\left(r,b\right)}=1/L{\left(r,b\right)}$.
}
The optimal non-zero \slope $p^{\tn{opt}}$ is then the real positive root of
\begin{equation}
\partial f/\partial p = A(1+p^2)^{3/2}-4p = 0,
\label{eq:cubic}
\end{equation}
where ${A = \left(r^{*}\right)^{2}/B^{*} \sqrt{1-(r^*)^2}}$, with ${B^{*}=B/(\mu R_0)}$ the ratio of bending stiffness to kink energy and ${r^*=r/R_0}$.
Surprisingly, we find that the optimal \slope is a discontinuous function of the helix radius (Fig.~\ref{fig:theory-pitch}):
for small $r$, \textit{i.e.}, $A < A_\tn{c} \equiv \sqrt{32/27}$, the optimal \slope is given by
\begin{equation}
    p^\tn{opt} = \sqrt{{\textstyle\frac{8}{A\sqrt{3}}}\cos{\left[{\textstyle\frac{1}{3}}\arccos{\left(-A\sqrt{{\textstyle\frac{27}{64}}}\right)}\right]} - 1}\;,
\label{eq:optpitch}
\end{equation}
whereas $p^\tn{opt}=0$ for $A > A_\tn{c}$.
At $A=A_\tn{c}$ the optimal \slope jumps from zero to $p^{\tn{opt}}=\sqrt{2}$, which, interestingly, describes a helix with an angle complementary to the magic angle $\theta_m=\arctan{\sqrt{2}}$.
This observation implies that the outer shells form a spool ($p=0$), while the inner shells form a twisted nematic domain with a large \slope (Fig.~\ref{fig:theory-pitch}A).

Whether the spool or the nematic core forms at low packing density depends on the ratio $B^{*}$.
$B^{*}>1$ favors the spool phase, whereas at $B^{*}<1$ the nematic core is more stable (Fig.~\ref{fig:theory-pitch}A, inset).
For a typical virus ($2R_0\approx50$~nm) and DNA parameters ($B=50 k_\tn{B}T\,\tn{nm},\, \mu=13.8k_\tn{B}T$) we find $B^{*}\approx0.14$ and so predict that a spool forms initially with a transition to a nematic core at radius $r_\tn{c}\approx 0.38R_0$.
Hence, the majority of the DNA should form a spool, enclosing a twisted-nematic core.

This result explains the experimental observation of a helical core structure in a dsRNA bacteriophage~\cite{ilca2019multiple}; taking $R_{0}=20.8\,\tn{nm}$ and $B=63k_\tn{B}T\,\tn{nm}$~\cite{abels2005single,hyeon2006size}, and assuming $\mu=13.8k_\tn{B}T$ (the value for dsDNA), we find excellent agreement with experiment on both the critical radius $r_\tn{c}$ and the helix angle at the transition~(Fig.\,\ref{fig:theory-pitch}A).
The analytical prediction of the helix angle also agrees with simulation results at low packing rates ($\omega\le0.0005\sigma\tau^{-1}$, Fig.~\ref{fig:theory-pitch}B), suggesting that, in the slow packing regime, our simulations are under thermodynamic control, and the packing morphology is determined by equilibrium free-energy minimization.

To obtain the full phase diagram of packing configurations we use the continuum theory of polymer nematics~\cite{Grason2,Svensek,curk2019} where the polymer is described by a vector field $\mathbf{t}(\mathbf{r})$.
Assuming that kinks form only at the enclosure surface, and disregarding entropic contributions due to conformational fluctuations, the lowest-order non-trivial terms of the elastic free-energy functional are given by,
\begin{equation}
E = K {\int_V} |\mathbf{t}| ~\Big(\hat{\mathbf{t}} \times (\nabla \times \hat{\mathbf{t}})\Big)^2~\tn{d}{\bf r} + \varepsilon_\textnormal{kink} {\oint_S}  |\mathbf{n} \cdot \mathbf{t}|~\tn{d}S   \;.
\label{eq-Ev}
\end{equation}
The first term is the bending term~\cite{Svensek,Grason2} with $\hat{\mathbf{t}}= \mathbf{t}/|\mathbf{t}|$ the unit tangent vector to the chain contour, while the second describes the kink density at the enclosure surface with normal vector $\mathbf{n}$.
This expression is an integral form of Eq.~\eqref{eq-Eel2} with prefactors $K=2B/{(\pi \sigma^2)}$ and $\varepsilon_\textnormal{kink}=2\mu/(\pi \sigma^2)$.

\begin{figure}[bt!]
\centering
\includegraphics[width=\columnwidth]{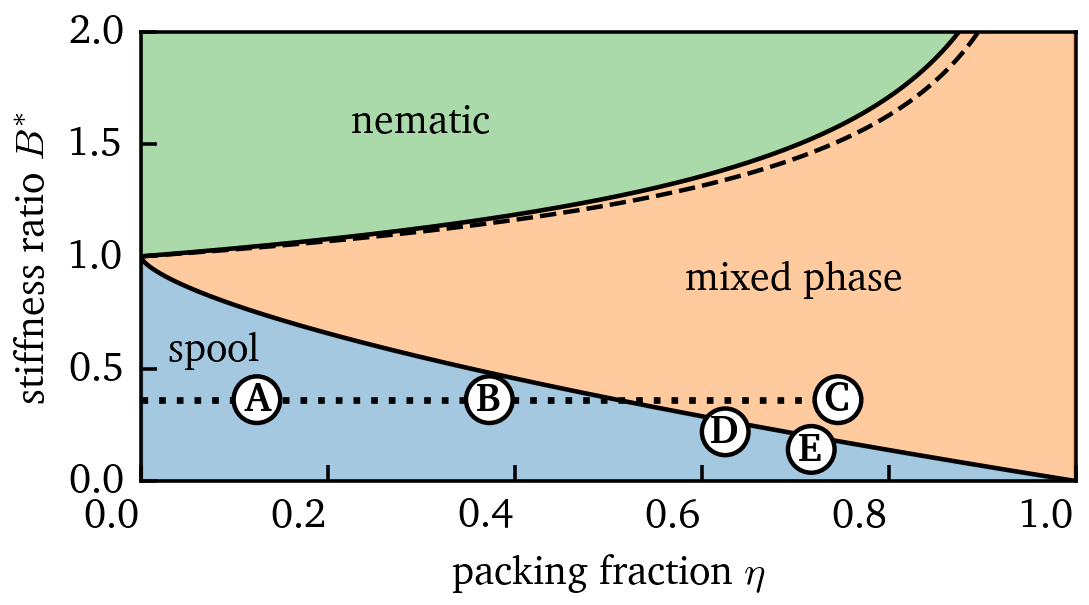}
\caption{Theoretical phase diagram showing the stable phases.
The phase boundaries (solid lines) are obtained by minimizing the total free energy $E$ at zero nematic twist [Eqs.~\eqref{eq:Es}, \eqref{eq:En}], whereas the dashed line shows boundary at optimal twist [Eq.~\eqref{eq:optpitch}] (the spool--mixed-phase boundaries overlap).
The dotted line indicates the range of parameter space probed by simulations.
Points A--C refer to the snapshots in Fig.~\ref{fig:order};
points D--F are estimates for bacteriophages $\phi6$~\cite{ilca2019multiple} and $\phi22$~\cite{tang2011peering}\editRevOne{, and herpes simplex virus type 1~\cite{Chen2023}}
at $\mu=13.8k_{\tn{B}}T$ (see \edit{\cite{Note1}} for details of their computation).
}
\label{fig:pdmix}
\end{figure}

The optimal structure of a confined meltable filament results from competition between the elastic and melting terms in Eq.~\eqref{eq-Ev}.
For stiff polymers, kinking is preferable to bending, so they will tend to arrange into nematically-aligned straight segments with kinks at the boundaries. 
In contrast, flexible polymers are expected to form spool structures without kinks.
To explore these competing mechanisms, we assume that the chains form a two-domain structure: an outer spool and a nematic core.
The spool contains no kinks ($\mathbf{n} \cdot \mathbf{t}=0$), while the nematic core does not contribute to bending ($\nabla \times \hat{\mathbf{t}}=0$).
Assuming phases are close packed, the spool free-energy is~\cite{curk2019,Purohit2003},
\begin{equation}
E_\tn{spl} (\eta, R_0) =  \frac{\pi^2 B R_0}{\sqrt{3}\sigma^2} \left[\textstyle{\frac{1}{2}}\ln  \frac{1+\eta^{1/3} }{1-\eta^{1/3} } - \eta^{1/3}\right]\;,
\label{eq:Es}
\end{equation}
while the nematic free energy is
\begin{equation}
E_\tn{nem} (\eta, R_0) = \frac{\pi^2 \mu R_0^2}{2\sqrt{3}\sigma^2}  \left[1-(1-\eta)^{2/3} \right]\;.
\label{eq:En}
\end{equation}
Since the spool and the nematic core occupy disjoint volumes, and connections between domains can be neglected in the long-chain limit, the total free energy is the sum,
 $E(\eta,\eta_{\tn{spl}},R_0)=E_\tn{spl} (\eta_{\tn{spl}}, R_0) + E_\tn{nem} (\eta-\eta_{\tn{spl}}, R_0)$, with $\eta_{\tn{spl}}$ the spool volume.
Minimizing $E(\eta,\eta_{\tn{spl}},R_0)$ we find a phase \mbox{diagram (Fig.~\ref{fig:pdmix})} that is a function of two dimensionless parameters:
$B^{*}=B/(\mu R_0)$ and $\eta$.
The boundary delineating the pure nematic and the mixed phase is $B^{*} = (1-\eta)^{-1/3}$, while the boundary between the pure spool and the mixed phase is $B^{*}=1-\eta^{2/3}$. 

The coexistence region emerges at $B^*\sim 1$, growing with $\eta$ and dominating in the high packing fraction limit. 
For a typical $50\,\tn{nm}$ virus, $B^{*}\approx 0.14$, implying a spool phase at low packing fractions, enclosing a nematic core at ${\eta>(1-B^{*})^{3/2}\approx0.8}$.
This theory assumes coupled shells with no nematic twist in the core ($p\to \infty$); considering the opposite limit of free shells slightly stabilizes the nematic phase by allowing non-zero twist [Eq.~\eqref{eq:optpitch}], but the shift in the phase boundaries is less than $0.1B^*$ (dashed line in Fig.~\ref{fig:pdmix}).

\editRevOne{Experimental data can be mapped onto the phase diagram by assuming that $\mu$ for RNA and DNA are the same, and the range of $\eta$ is bounded by hexagonal and cubic packings (both observed experimentally). 
With such a mapping, the observed spool--twisted-nematic structure packings in $\phi6$ bacteriophage~\cite{ilca2019multiple} and herpes simplex virus~\cite{Chen2023} are consistent with our theoretical predictions (Fig.~\ref{fig:pdmix}).
}

In summary, our theoretical and simulation results imply that \mbox{dsDNA/dsRNA} packed into a sphere adopts a spool--nematic configuration.
Coexistence between an outer spool and a nematic core arises spontaneously both in packing simulations and equilibrium theory based on the minimization of elastic energy.
We explored the strong and weak limits of orientational coupling, finding nearly-identical phase diagrams for a meltable chain, with the main difference being whether the nematic core exhibits non-zero twist.
We expect that packing of \mbox{dsDNA} and \mbox{dsRNA} in viruses falls between these two limits, forming a large outer spool with a (twisted-)nematic core emerging at high packing densities ($\eta>0.8$ for a typical $2R_0=50$ nm virus).
These results explain the experimental data for dsRNA packing in a bacteriophage virus~\cite{ilca2019multiple}.

We have considered packing into a pre-formed spherical capsid, but our findings are based on global free-energy minimization and thus expected to hold qualitatively for other packing situations such as DNA condensation or ordering in bacterial nucleomes.
For example, kink formation and nematic ordering may explain why plasmid DNA assembles into rod-like structures~\cite{xuan2013plasmid} rather than tori~\cite{hoang2014fromtoroidal}.
\edit{Moreover, our phase diagram shows that spool--nematic ordering is expected for any strongly confined kinkable polymer. 
Non-linear bending and kinking behavior might be a common characteristic of helical polymers, as their structure is usually determined by reversible supramolecular interactions.
}

\begin{acknowledgments}
This work was supported by the Whiting School of Engineering (JHU) through startup funds,  
the Chinese National Science Foundation through the Key research grant 12034019, and from the Strategic Priority Research Program of the Chinese Academy of Sciences through the grant XDB33000000.
\end{acknowledgments}


%

\end{document}